\documentclass[11pt,letterpaper]{article}

\usepackage{amsmath,amsthm}
\usepackage{mathtools}
\mathtoolsset{showonlyrefs}

\usepackage{ifluatex,ifxetex}
\ifxetex
\usepackage{xltxtra}
\else
\ifluatex
\usepackage{luatextra}
\else
\usepackage{amssymb}
\usepackage[utf8]{inputenc}
\usepackage[T1]{fontenc}
\usepackage{mathpazo,tgcursor,tgheros,tgpagella}
\usepackage{textcomp}
\usepackage[mathscr]{eucal}
\providecommand{\mathbfscr}[1]{\boldsymbol{\mathscr{#1}}}

\csname else\expandafter\endcsname
\romannumeral -`0
\fi
\fi
\iftrue\relax%
\defaultfontfeatures{Ligatures=TeX}
\usepackage[math-style=TeX, vargreek-shape=TeX]{unicode-math}
\AtBeginDocument{%
  \let\setminus\smallsetminus
}
\fi

\usepackage[margin=1in]{geometry}

\usepackage{graphicx,color,authblk}
\usepackage[authoryear]{natbib}

\usepackage{paralist}
\usepackage{enumitem}
\newlist{enumerate*}{enumerate*}{1}
\setlist[enumerate*]{label=(\arabic*),
  itemjoin={{, }}, itemjoin*={{, and }}}

\usepackage{xparse}

\usepackage[bookmarksnumbered,bookmarksopen,
unicode]{hyperref}

\hypersetup{
  pdftitle={The matching problem does not admit fully-polynomial size
  linear programming relaxation schemes},
  pdfauthor={Gábor Braun,
    Sebastian Pokutta},
  pdfkeywords={approximate polyhedral complexity,
    matching polytope,
    common information as lower bound on nonnegative rank},
  pdfsubject={MSC2000 Primary
    52B12, 
    68Q17; 
    Secondary
    05C70, 
    94A17 
  }} 

\DeclareMathOperator{\nnegrk}{rk_+} 

\DeclareMathOperator{\poly}{poly} 
\newcommand{\R}{\mathbb{R}}
\newcommand{\N}{\mathbb{N}}

\newcommand{\match}[1]{P_M(#1)}
\newcommand{\pmatch}[1]{P_{PM}(#1)}

\DeclareMathOperator{\fc}{fc}
\DeclareMathOperator{\oprob}{\mathcal{P}}
\newcommand{\problem}[1]{\textnormal{\textsf{#1}}}
\newcommand{\matching}[1]{\problem{Matching(\ensuremath{#1})}}

\newcommand{\face}[1]{\left\{#1\right\}}
\newcommand{\card}[1]{\left|#1\right|}

\newcommand{\binSet}{\{0,1\}}

\newcommand*{\set}[2]{\left\{#1\,\middle|\,#2\right\}}

\NewDocumentCommand{\conv}{mo}{\operatorname{conv}\left(#1%
  \IfValueT{#2}{\,\middle|\,#2}\right)}

\DeclarePairedDelimiter{\abs}{|}{|}

\newcommand*{\VAR}[1]{\mathbf{#1}}
\newcommand*{\EVENT}[1]{\mathscr{#1}}
\newcommand*{\VAREVENT}[1]{\mathbfscr{#1}}

\NewDocumentCommand{\probability}{d()om}{%
  \operatorname{\mathbb{P}}%
  \IfValueT{#1}{\sb{#1}}%
  \left[#3%
    \IfValueT{#2}{\,\middle|\,#2}\right]}
\NewDocumentCommand{\expectation}{d()om}%
  {\operatorname{\mathbb{E}}%
      \IfValueT{#1}{\sb{#1}}%
      \left[#3%
    \IfValueT{#2}{\,\middle|\,#2}\right]}
\NewDocumentCommand{\entropy}{om}{\mathbb{H}\left[#2
    \IfValueT{#1}{\,\middle|\,#1}\right]}
\NewDocumentCommand{\bentropy}{lm}
  {\widetilde{\mathbb{H}}#1\left[#2\right]}
\newcommand*{\relEntropy}[2]
  {\operatorname{\mathbb{D}}\left(#1\,\middle\|\,#2\right)}
\NewDocumentCommand{\mutualInfo}{omm}{\mathbb{I}\left[#2;#3
    \IfValueT{#1}{\,\middle|\,#1}\right]}
\NewDocumentCommand{\privInfo}{omm}{\mathbb{W}\left[#2;#3
    \IfValueT{#1}{\,\middle|\,#1}\right]}
\DeclareExpandableDocumentCommand{\comm}{om}
  {\mathbb{C}\left[#2\IfValueT{#1}{\,\middle|\,#1}\right]}
\NewDocumentCommand{\priv}{om}
  {\mathbb{W}\left[#2\IfValueT{#1}{\,\middle|\,#1}\right]}
\NewDocumentCommand{\commonInfo}{omm}{\mathbb C\left[#2;#3
    \IfValueT{#1}{\,\middle|\,#1}\right]}

\NewDocumentCommand{\infrk}{om}
  {\mathbb{C}\left[#2\IfValueT{#1}{\,\middle|\,#1}\right]}

\newcommand*{\size}[1]{\left|#1\right|}

\DeclareMathOperator{\range}{range}

\newcommand*{\Mbad}[1]{\operatorname{\VAR M-BAD}(#1)}
\newcommand*{\Ubad}[1]{\operatorname{\VAR U-BAD}(#1)}
\newcommand*{\bad}[1]{\operatorname{BAD}(#1)}
\newcommand*{\good}[1]{\operatorname{GOOD}(#1)}

\newtheorem{theorem}{Theorem}[section]
\newtheorem{lemma}[theorem]{Lemma}
\newtheorem{proposition}[theorem]{Proposition}

\theoremstyle{definition}

\newtheorem{defn}[theorem]{Definition}
\newtheorem{definition}[theorem]{Definition}

\newtheorem{exa}[theorem]{Example}

\theoremstyle{remark}

\newtheoremstyle{claim}{}{}{}{}{\itshape}{.}{.5em}{}
\theoremstyle{claim}

  {\end{proof}}

\title{The matching problem has no fully-polynomial size \\
  linear programming relaxation schemes\footnote{This paper was presented in part at the Symposium on Discrete Algorithms (SODA).}}

\author{Gábor Braun and Sebastian Pokutta \\
  ISyE,
   Georgia Institute of Technology, \\
   Atlanta, GA 30332, USA. \\
   \textit{Email:}~gabor.braun@isye.gatech.edu,
   sebastian.pokutta@isye.gatech.edu}
\date{}

\begin{document}
\maketitle

\begin{abstract}
  The groundbreaking work of \cite{rothvoss2013matching} established
  that every linear program expressing the matching polytope has an
  exponential number of inequalities
  (formally,
  the matching polytope has exponential extension complexity). We
  generalize this result by deriving strong bounds on
  the LP inapproximability
  of the matching problem:
  for fixed \(0 < \varepsilon < 1\),
  every \((1 - \varepsilon / n)\)-approximating LP requires
  an exponential number of inequalities,
  where \(n\) is the number of vertices.
  This is sharp given the well-known 
  \(\rho\)-approximation of size \(O(\binom{n}{1/(1 - \rho)})\)
  provided by the odd-sets of size up to \(1/(1 - \rho)\).
  Thus matching is the first problem in \(P\),
  which does not
  admit a \emph{fully polynomial-size LP relaxation scheme}
  (the LP equivalent of an FPTAS),
  which provides a sharp separation from the
  \emph{polynomial-size LP relaxation scheme}
  obtained e.g., via constant-sized
  odd-sets mentioned above.

  Analyzing the size of LP formulations is equivalent to examining the
  nonnegative rank of matrices.  We study the nonnegative rank
  via an information-theoretic approach, and while it
  reuses key ideas from \cite{rothvoss2013matching}, the main
  lower bounding technique is different: we employ the information-theoretic notion of \emph{common
    information} as introduced in \cite{wyner1975common} and used for
  studying LP formulations in \cite{BP2013commInfo}. This allows us to
  analyze the nonnegative rank of perturbations of slack matrices,
  e.g., approximations of the matching polytope.
  It turns out that the high
  extension complexity for the matching problem stems from the
  same source of hardness as in the case of the correlation polytope:
  a direct sum structure.
\end{abstract}

\section{Introduction}
\label{sec:introduction}
In recent years, there has been a significant interest in
understanding the expressive power of linear programs. The motivating
question is which polytopes or combinatorial optimization problems can
be expressed as linear programs with a small number of
inequalities. The theory of \emph{extended formulations} studies  this question and the measure of interest is the \emph{extension
  complexity} which is the size of the smallest
linear programming formulation (in terms of the number of inequalities).
This notion of complexity is independent of P vs.~NP and
is characterized by nonnegative matrix factorizations. We will
analyze this factorization problem
for the all-important matching
problem using information theory.

Most of the recent progress in extended formulations is ultimately rooted in \cite{Yannakakis88,Yannakakis91}
seminal paper ruling out \emph{symmetric} linear programs with a
polynomial number of inequalities for the TSP polytope and the
matching polytope.
It was the innocent question whether the
same still holds true when the symmetry assumption is removed
that spurred a significant body of work (see
Section~\ref{sec:hist-extend-form} below).
Yannakakis's question was answered in the affirmative
in \cite{extform4} for the TSP polytope via the correlation polytope,
and in \cite{rothvoss2013matching}
for the matching polytope.

The fact that the matching problem has high
extension complexity raises some fundamental questions. After all,
matching can be solved in polynomial time. Two key questions are
\begin{enumerate*}
\item
  how well we can approximate the matching problem
  via a linear program
\item
  what
commonality the correlation polytope and the matching problem possess
so that both require an exponential number of inequalities.
\end{enumerate*}
In this work, we answer both of these questions. 

We generalize the results in \cite{rothvoss2013matching}
to show in Theorem~\ref{th:approx-matching}
that the matching problem cannot be approximated within a
factor of \(1 - \Theta(1/n)\) by a linear program with a
polynomial number of inequalities.
Thus, the matching problem does not admit a
fully polynomial-size linear programming relaxation scheme (FPSRS),
the
linear programming equivalent of an FPTAS, i.e., it cannot be
\((1 - \varepsilon)\)-approximated with a linear program of size
\(\poly(n,1/\varepsilon)\).
This complements the folklore polynomial-sized
\((1 - 1/k)\)-approximate linear programming formulation for the
matching problem (see Example~\ref{ex:PSRS-matching}),
establishing a threshold of \(\Theta(1/n)\)
between polynomial-size approximability
and exponential inapproximability.

Further, our approach is based on a direct information-theoretic argument
extending the techniques developed in \cite{BP2013commInfo}. 
This proof exhibits key commonalities with the
one for the correlation polytope, revealing
that the reason for high complexity of both, the matching problem and
the correlation polytope, is an (almost identical) direct sum
structure contained in both problems.

\subsection{History of extended formulations}
\label{sec:hist-extend-form}

We will now provide a brief overview of extended formulations.  As mentioned before,
the interest in extended formulation was initiated by
the question of Yannakakis whether the TSP polytope and the matching polytope
have (asymmetric) linear programs with a polynomial number
of inequalities.
In \cite{KaibelPashkovichTheis10} it was shown that the symmetry actually
can make a significant difference for the size of extended
formulations. The authors studied the \(\ell\)-matching polytope
(polytope of all matchings with exactly \(\ell\) edges),
closely related to Yannakakis's work prompting the question whether
every 0/1 polytope has an efficient
polyhedral lift (note that this is independent of P vs. NP
as we disregard encoding length). This question was answered in the
negative in \cite{Rothvoss11}, establishing the existence of 0/1
polytopes requiring an exponential number of inequalities
in any of their formulations, and it was recently extended to
the case of SDP-based extended formulations in
\cite{BDP2013}. The same technique was used in
\cite{FioriniRothvossTiwary11} to derive lower bounds on the extension
complexity of polygons in the plane,
which are of prototypical importance
for many related applications; the analogous results for the SDP case
have been obtained in \cite{BDP2013}. 

Shortly after Rothvoß's result it was shown in
\cite{extform4} that the first half of Yannakakis's question is in the
affirmative: the TSP polytope requires an exponential number of linear
inequalities in any linear programming formulation, irrespective of P
vs.~NP. In fact, it was shown that the correlation polytope (or
equivalently the cut polytope) has extension complexity \(2^{\Omega(n)}\)
where \(n\) is the length of the bit strings, and
the stable set polytope (over a
certain family of graphs) as well as the TSP polytope have extension
complexity \(2^{\Omega(n^{1/2})}\) where \(n\) is the number of nodes
in the graph. In \cite{bfps2012} the result for
the correlation polytope was generalized to the approximate case,
showing that the CLIQUE polytope,
also called the correlation polytope,
cannot be approximated with
a polynomial-size linear program better than \(n^{1/2-\varepsilon}\),
which was subsequently improved in \cite{braverman2012information} to
\(n^{1-\varepsilon}\) matching Håstad's celebrated inapproximability result for
the CLIQUE problem (see \cite{haastad1999clique}). In
\cite{BP2013commInfo} the results for the CLIQUE polytope was
further improved to average-case type results via
a general
information-theoretic framework to lower bound the nonnegative rank of
matrices. In \cite{BFP2013} these average-case arguments were used to
show that the average-case (as well as high-probability)
extension complexity of the stable-set problem
is high. Very recently in \cite{LRS14} strong explicit lower bounds on the semidefinite extension complexity of the correlation polytope (and many other problems) have been obtained. 

Additional bounds for various polytopes including the
knapsack polytope have been established in \cite{VP2013} and
\cite{2013arXiv1302.2340A} using the reduction mechanism outlined in
\cite{extform4}.
Besides CLIQUE and stable set,
there are only few results on approximability of problems by
linear programs.
For any fixed \(\varepsilon > 0\),
a polynomial-sized linear program giving a
\((1 - \varepsilon)\)-approximation of the knapsack problem
has been provided in \cite{bienstock2008approximate}.
An
exponential lower bound on the approximate polyhedral complexity of
the metric capacitated facility location problem  was established
in \cite{KolliopoulosM13a},
however only in the original space;
extended formulations were not considered. 

Recently, it has been
observed that the same techniques (and in fact essentially identical
proofs) can be used to obtain results on the size of formulations
independent of the actual encoding of the problem as polytope. This
generalization was first observed for uniform formulations  in
\cite{chan2013approximate}, extended to general (potentially non-uniform)
problems in \cite{BFP2013} for analyzing the average case complexity
of the stable set problem, and fully generalized to affine functions
allowing for approximations and reductions in
\cite{BSZ2014._lp_sdp} providing inapproximability results for
various problems, such as e.g., vertex cover. All these versions are
extensions of the initial approach to approximations
via polyhedral pairs
(see \cite{bfps2012,Pashkovich12}). 

The main tool for lower bounds
is Yannakakis's celebrated Factorization Theorem stating
that the extension complexity
of a polytope is equal to the nonnegative rank of any of its
\emph{slack matrices},
which has been extended to the approaches mentioned above.
Combinatorial as well as communication based lower bounds
for the nonnegative rank have been explored in
\cite{FaenzaFioriniGrappeTiwary11,FioriniKaibelPashkovichTheis11}.
At the core of all of the above super-polynomial lower bounds
on the extension complexity is the all-important UDISJ problem whose
partial matrix appears as a pattern in the slack matrix of these
problems.
The current best lower bound on the nonnegative rank of the UDISJ
matrix is established in \cite{kaibel2013short}
by a remarkably short combinatorial argument.

Two notable exceptions not using UDISJ are \cite{chan2013approximate,LRS14},
mentioned above adapting Sherali-Adams and Lasserre separators to obtain lower bounds.
Another exception is the very recent result of
\cite{rothvoss2013matching}, which finally answers the second half of
Yannakakis's question, showing that the matching polytope has
exponential extension complexity. The proof is based on Razborov's
technique (see \cite{Razborov92}),
and provides the second base matrix with linear rank but exponential
nonnegative rank, namely,
the slack matrix of the matching problem.

\subsection{Related work}
\label{sec:related-work} 

We recall the works whose methodology is closely related to ours.
The most closely related one is \cite{BP2013commInfo}
providing a general information-theoretic framework for lower bounding the extension
complexity of polytopes in terms of information, motivated by
\cite{braverman2012information3} as well as
\cite{jain2013efficient}. A dual information-theoretic approach, similar to the one for the
fractional rectangle covering number (see e.g., \cite{KarchmerKushilevitzNisan95}) has been explored in
\cite{BJLP2013}, however, we will stick to the primal approach here,
dealing directly with distributions over potential rank-1 matrices. The
cornerstone of our framework is the notion of common information
introduced by \cite{wyner1975common} for (a completely unrelated)
use in information theory.  For estimating information,
\cite{BP2013commInfo} used Hellinger distance inspired by
\cite{BarYossef2004702}, however here it is more effective to employ
Pinsker's inequality.
The model of approximate linear programs is the same as in \cite{BSZ2014._lp_sdp} 
based on \cite{chan2013approximate}; see also \cite{BFP2013}.

\subsection{Contribution}
\label{sec:contribution}

Our main result is Theorem~\ref{th:approx-matching}
proving inapproximability of the matching polytope, providing the
following three main contributions.
We use the term \emph{formulation complexity} for the complexity of
linear programming formulations (again measured in the number of required linear inequalities),
as introduced in \cite{BSZ2014._lp_sdp}.
This notion is a natural generalization of extension complexity,
not requiring a polytope encoding for optimization problems.

\begin{enumerate}
\item \textbf{Polyhedral inapproximability of the matching problem:}
 As mentioned above,
 the matching problem can be solved in polynomial time,
 even though its formulation complexity
 is exponential.
 However, it can be \((1 - \varepsilon)\)-approximated
 by a linear program of polynomial size for an \emph{fixed}
 \(\varepsilon > 0\), i.e.,
 it admits a
 \emph{polynomial-size linear programming relaxation scheme (PSRS)},
 which is the linear programming equivalent of a PTAS. 
 As a consequence, an important and natural question is
 whether one can  even go a step further and find a linear
 program of size \(\poly(n,1/\varepsilon)\) approximating the matching
 polytope within a factor of \(1 - \varepsilon\),
 i.e., whether it admits a
 \emph{fully polynomial-size linear programming relaxation scheme
   (FPSRS)}; the linear programming equivalent of an FPTAS.
 We resolve this question in the
 negative, showing that the formulation complexity
 and (classical)
 computational complexity are very different notions. The
 non-existence of an FPSRS for matching can be also interpreted
 as an indicator that matching might be
 especially hard from a linear programming point of view,
 similar to the notion of strong
NP-hardness which also precludes the existence of an FPTAS (in most
cases) in the computational complexity setup. 

\item \textbf{Information-theoretic proof:} We provide an information-theoretic proof for the lower bound on
  the linear programming complexity of the matching problem and its
  approximations. The proof is based on
  an extension of the framework in  \cite{BP2013commInfo}. By arguing
  directly  via the distribution of potential factorizations, rather than considering single rank-1 factors or rectangles as
  typically required by Razborov's method, the
  setup can be considerably simplified and the information-theoretic
  framework naturally lends itself to approximations.
  We obtain a simple and short
  proof that provides additional insight into the structure of
  matchings and their linear programming hardness.

\item \textbf{High complexity implied by direct sum
    structure:}
  The information-theoretic approach is well suited
  for proving exponential lower bounds by partitioning
  the structure \(P\)
  into copies of a fixed-size substructures \(P_{0}\),
  and applying a direct sum argument to show
  \[\log \fc(P) \geq \Theta(n) \cdot \log \fc(P_0),\]
  where \(\fc(Q)\) denotes the formulation complexity of \(Q\)
  and the bound on \(\fc(P_0)\) is non-trivial.
  This was already used for the correlation polytope
  in \cite{BP2013commInfo}, and we show that the matching polytope
  exhibits a similar direct sum structure.
  The partition we use is a simplified variant of the one in 
  \cite{rothvoss2013matching}, however it is still slightly more
  involved 
  than the one for the correlation polytope
  due to additional structure we need to consider.
\end{enumerate}

  In the above direct sum framework, the difficulty arises from actually having
  an information-theoretic quantity in place of
  \(\log \fc(P_{0})\), which is used to make the direct sum argument
  work in first place. Thus a positive constant lower bound is not
  immediate, in particular when considering approximations.

The authors believe that the presented approach can be also used to obtain 
linear programming inapproximability results for a variety of other problems.

\subsection{Outline}
\label{sec:outline}
We start with preliminaries in Section~\ref{sec:preliminaries}, introducing
necessary notions and notations from information theory in
Section~\ref{sec:inform-theor-basics} as well as providing a recap of
the theory of (approximate) LP formulations in
Section~\ref{sec:formulation-complexity}.
In Section~\ref{sec:lower-bounds-nonn}
we provide the connection between common information and
nonnegative rank, which is subsequently extended to provide bounds for
the matching problem in Section~\ref{sec:lower-bounds-extens}.  We
choose notations mostly consistent with \cite{rothvoss2013matching}
for easy relation of both approaches. 

\section{Preliminaries}
\label{sec:preliminaries}

We use \([k]
\coloneqq \face{1,\dots,k}\) as a short-hand. 
Let \(\log\) denote the logarithm to base \(2\).
We will denote random variables by bold capital letters
such as \(\VAR{A}\) to avoid
confusion with sets that will also be denoted by capital
letters. Events will be denoted by capital script letters such as
\(\EVENT{E}\) when not written out,
but conditions
will be denoted by script bold letters such as
\(\VAREVENT Z\)
in conditional probabilities and other conditional quantities:
conditions can be events, random variables, or combinations
of both types.
We use \(\perp\) to indicate
independence: e.g., \(\VAR{A} \perp \VAR{B}\).

\subsection{Information-theoretic basics}
\label{sec:inform-theor-basics}

We briefly recall standard basic notions from information theory
in this section and we refer the reader to \cite{cover2006elements} for more details
and as an excellent introduction.
Information is measured in bits,
as is standard for discrete random variables.

We will now recall the notion of mutual information which is
at the core of our arguments. The
\emph{mutual information} of two discrete random variables
\(\VAR{A}, \VAR{B}\) is defined as
\[\mutualInfo{\VAR{A}}{\VAR{B}} \coloneqq
\sum_{\substack{a \in \range(\VAR A) \\ b \in \range(\VAR B)}}
\probability{\VAR{A} = a, \VAR{B} = b}
\log
\frac{\probability{\VAR{A} = a, \VAR{B} = b}}
{\probability{\VAR{A} = a} \cdot \probability{\VAR{B} = b}}
.\]
It captures how much information about \(\VAR A\)
is leaked by considering \(\VAR B\); and vice versa:
mutual information is symmetric.
We will often
have \(\VAR A\) and \(\VAR B\) being  a collection of random
variables. We use a comma to separate the
components of \(\VAR A\) or \(\VAR B\), and a semicolon to separate
\(\VAR A\) and \(\VAR B\) themselves, e.g., 
\(\mutualInfo{\VAR A_{1}, \VAR A_{2}}{\VAR B} = \mutualInfo{(\VAR
  A_{1}, \VAR A_{2})}{\VAR B}\). We can naturally extend
mutual information to \emph{conditional mutual information}
\[\mutualInfo[\VAR C, \EVENT E]{\VAR A}{\VAR B}
\coloneqq \expectation(\VAR{C} \mid \EVENT{E})
{\mutualInfo{(\VAR{A} \mid \VAR{C}, \EVENT{E})}
{(\VAR{B} \mid \VAR{C}, \EVENT{E})}}
\]
by using the respective conditional distributions,
where \(\VAR C\) is a
random variable and \(\EVENT E\) is an event.
Note that the expectation is implicitly taken over random variables
in the condition.

We shall use the following bounds on mutual information.
An obvious upper bound is
\begin{equation*}
  \mutualInfo[\VAREVENT{Z}]{\VAR{A}}{\VAR{B}}
  \leq \log \size{\range(A)},
\end{equation*}
which provides the lower bound
on the logarithm of the nonnegative rank
in Lemma~\ref{lem:nonnegrank-via-IC}.

Exponential lower bounds of the nonnegative rank will be obtained via
the \emph{direct sum property}, which states that
for mutually independent random variables
\(\VAR{A_{1}}, \dotsc, \VAR{A_{n}}\)
given a condition \(\VAREVENT{Z}\) we have 
\begin{equation*}
  \mutualInfo[\VAREVENT{Z}]{\VAR{A_{1}}, \dotsc, \VAR{A_{n}}}
  {\VAR{B}}
  \geq
  \sum_{i \in [n]}
  \mutualInfo[\VAREVENT{Z}]{\VAR{A_{i}}}{\VAR{B}}
  .
\end{equation*}
In order to lower bound each summand we will use a divergence measure.
The \emph{relative entropy} or \emph{Kullback–Leibler divergence}
measures the difference of two probability distributions.
It is always nonnegative, but it is neither symmetric,
nor does it satisfy the triangle inequality.
For simplicity we only define it for random variables.

\begin{definition}[Relative entropy] Let \(\VAR X,\VAR Y\) be discrete random
  variables on the same domain.
  \emph{Relative entropy of \(\VAR{X}\) and \(\VAR{Y}\)} is
  \[\relEntropy{\VAR{X}}{\VAR{Y}} =
  \sum_{x \in \range(\VAR{X})}
  \probability{\VAR X = x} \cdot \log
  \frac{\probability{\VAR X = x}}{\probability{\VAR Y = x}}.\]
  Here \(0 \log (0/q) = 0\) and \(p \log (p/0) = + \infty\)
  for \(p > 0\) by convention.
\end{definition}

Relative entropy is related to mutual information via the
following identity:
\[\mutualInfo{\VAR X}{\VAR Y} = \expectation(\VAR Y){\relEntropy{\VAR X\mid \VAR Y}{\VAR X}},\]
i.e., mutual information is the expectation of the deviation over
\(\VAR Y\).  Pinsker's inequality provides a convenient lower bound on the
relative entropy (see e.g., \cite[Lemma 11.6.1]{cover2006elements}).
\begin{lemma}[Pinsker's inequality]
\label{lem:pinsker}
  Let \(\VAR X,\VAR Y\) be discrete random
  variables with identical domains. Then
\[\relEntropy{\VAR X}{\VAR Y} \geq 2 (\log e)
  \left(
    \max_{\EVENT{E} \colon \textup{ event}}
    \abs*{\probability{\EVENT{E}(\VAR{X})} -
    \probability{\EVENT{E}(\VAR{Y})}}
  \right)^{2}
.\]
\end{lemma}

The quantity
\(\max_{\EVENT{E}: \textup{ event}} \abs*{\probability{\EVENT{E}(\VAR{X})}
- \probability{\EVENT{E}(\VAR{Y})}}\)
is called
the \emph{total variation distance}
between \(\VAR{X}\) and \(\VAR{Y}\),
which is the maximal difference of probabilities
the distributions of \(\VAR{X}\) and \(\VAR{Y}\)
assign to the same event.
For mutual information, via Pinsker's inequality we obtain the lower bound
\[\mutualInfo{\VAR{X}}{\VAR{Y}} \geq 2 (\log e) \cdot
  \expectation(\VAR{Y})
  {\max_{\EVENT{E}: \textup{ event}}
    \abs*{\probability{\EVENT{E}(\VAR{X})} -
    \probability[\VAR{Y}]{\EVENT{E}(\VAR{X})}}^{2}}
.\]

\subsection{Approximations via linear programs}
\label{sec:formulation-complexity}

We will now recall the framework of linear programming formulations
from \cite{BSZ2014._lp_sdp}, subsuming previous approaches via
\emph{polyhedral pairs},
but with the advantage of requiring no linear encoding
for the problem statement. It also includes a reduction to matrix factorizations,
reducing bounds on the size of linear programming formulations
to bounds of nonnegative rank of matrices.

We refer the interested reader to the excellent surveys
\cite{ConfortiCornuejolsZambelli10} and \cite{Kaibel11} as well as
\cite{Pashkovich12,bfps2012} for the approximate extended formulation
framework via polyhedral pairs.

We are interested in questions of the following type:
\begin{quote}
  \begin{center}
  \emph{How hard is it to approximate a maximization problem
    \(\oprob\) within \\ a factor of \(\rho\) via a linear programming formulation?}    
  \end{center}
\end{quote}
One could ask the same question for minimization problems,
and the framework below also applies to them,
but for simplicity, here we restrict the exposition to maximization problems only.

We start by making the notion of a maximization problem precise.
A \emph{maximization problem}
\(\oprob = (\mathcal{S}, \mathcal{F})\)
consists of a set of feasible solutions \(\mathcal{S}\)
and a set of objective functions \(\mathcal{F}\)
defined on \(\mathcal{S}\).
The aim is to approximate the maximum value of every objective
function in $\mathcal{F}$ over $\mathcal{S}$. An \emph{approximation problem}
\(\oprob^{*} = (\oprob, \mathcal{F}^{*})\)
is a maximization problem \(\oprob\)
together with a family
\(\mathcal{F}^{*} = \set{f^{*}}{f \in \mathcal{F}}\)
of numbers called \emph{approximation guarantees} satisfying
\(\max f \leq f^{*}\).
An algorithm solves \(\oprob^{*}\)
if it computes for every \(f \in \mathcal{F}\)
an approximate value \(\widehat{f}\)
satisfying \(\max f \leq \widehat{f} \leq f^{*}\).
This is a formalization of the traditional notion of approximation
algorithms.  For example setting \(f^{*} = \max f\)
we compute the exact maximum values.  Often the objective functions
are nonnegative, and one wants to approximate the maximum within a
factor \(0 < \rho < 1\).
Recall that an approximation factor \(\rho\)
for an approximation algorithm means that the algorithm computes a
solution \(s\)
with \(f(s) \geq \rho \max f\), and in order to be compatible with
this model we use the equivalent outer approximation in our model,
i.e., we choose \(f^{*} = \max f / \rho\).

We will consider the matching problem with the following
specification.

\begin{definition}[Matching problem]
  The matching problem \(\matching{n}\)
  has as feasible solutions all matchings \(M\) of
  the complete graph \(K_{n}\) on \([n]\).
  The objective functions \(f_{G}\) are indexed by all graphs
  \(G\) with \(V(G) \subseteq [n]\)
  with function values
  \begin{equation*}
    f_{G}(M) \coloneqq \size{M \cap E(G)}.
  \end{equation*}
\end{definition}
Note that \(M \cap E(G)\) is a matching of \(G\),
and a matching \(M\) of \(G\) is also a matching of \(K_{n}\).
In particular, \(\max f_{G} = \mu(G)\) is the \emph{matching number} of \(G\).

\subsubsection{LP formulation}
\label{sec:lp-formulation}

To complete the  framework, we will now define
linear programming formulations.  Given an approximation problem
\(\oprob^{*} = (\oprob, \mathcal{F}^{*})\) with \(\oprob =
(\mathcal{S}, \mathcal{F})\) as above, an \emph{LP formulation} of
\(\oprob^*\) consists of
\begin{enumerate}
\item
  a linear program \(A x \leq b\)
  with variables \(x \in \R^{d}\)
  for some \(d\);
\item
  a realization \(x^{s} \in \R^{d}\)
  for every feasible solution \(s \in \mathcal{S}\)
  satisfying \(A x^{s} \leq b\);
\item
  a realization \(w^{f} \colon \R^{d} \to \R\)
  for every objective function \(f \in \mathcal{F}\)
  as an affine map
  satisfying \(w^{f}(x^{s}) =f(s)\),
\end{enumerate}
and we require that the linear program
maximizing \(w^{f}(x)\) subject to \(A x \leq b\)
yields a solution within the approximation guarantee \(f^{*}\),
i.e., \(\widehat{f} \coloneqq \max_{x: A x \leq b} w^{f}(x) \leq f^{*}\).

The \emph{size} of an LP formulation is the number of inequalities
in \(A x \leq b\) and the \emph{formulation complexity} \(\fc(\oprob^{*})\)
of \(\oprob^{*}\) is the minimum size of its LP formulations.
When approximation guarantees are given
via an approximation factor \(0 < \rho \leq 1\),
i.e., \(f^{*} = \max f / \rho\), we shall write \(\fc(\oprob, \rho)\)
for the formulation complexity.

The matching polytope  provides an LP formulation
of the exact matching problem:

\begin{exa}[The matching polytope as formulation of the matching problem]

The \emph{matching polytope \(\match{n}\)} is defined as
the convex hull of the characteristic vectors \(\chi_{M}\)
of all matchings \(M\):
\[\match{n} \coloneqq \conv{\chi_M \in \R^{\binom{n}{2}}}%
[M \text{ is a matching in } K_n].\]
Recall from \cite{Edmonds65} that
the matching polytope \(\pmatch{n}\)
is the solution set of the linear program
\begin{align*}
  x(\delta(v)) &\leq 1 & v &\in [n], \\
  x(E[U]) &\leq \frac{\card{U} - 1}{2}
  & U &\subseteq [n], \card{U} \text{ odd}, \\
  x &\geq 0.
\end{align*}
Here and below for a vertex set \(U \subseteq [n]\),
let \(E[U]\) denote the set of edges
of the graph \(K_n\) contained in
\(U\), and let \(\delta(U)\) denote
the set of edges with one endpoint in \(U\) and one in its
complement; we use \(\delta(v) \coloneqq \card{\delta(\face{v})}\)
to denote the degree of \(v
\in [n]\).
Finally, \(x(E) \coloneqq \sum_{\{i,j\} \in E} x_{i,j}\) denotes
the sum of the coordinates of \(x\) from an arbitrary edge set \(E\).

This linear program provides an LP formulation for \(\matching{n}\)
with realizations \(x^{M} = \chi_{M}\)
for matchings \(M\) and the linear functions \(w^{f_{G}}(x) = x(E(G))\)
for objective functions \(f_{G}\). Observe that most of the inequalities
have the form \(w^{f_{K_{U}}}(x) \leq \max f_{K_{U}}\), which will be
useful later.
\end{exa}

\subsubsection{Polynomial-size linear programming relaxation
  schemes}
\label{sec:polyn-size-relax}

We will also examine the trade-off
between approximation factor and size.
The following two notions are the linear programming equivalents of
PTAS and FPTAS.

\begin{definition}[(Fully) polynomial-size linear programming
  relaxation scheme]
  Let \(\oprob_{n}\)
  with \(n \in \N\) be a family of maximization problems.
  The family \(\{\oprob_{n}\}_{n \in \N}\) admits a
  \begin{enumerate}
  \item \emph{polynomial-size linear programming
      relaxation scheme (PSRS)}
    if for every fixed
    \(\varepsilon > 0\)
    there is an LP formulation of \(\oprob_{n}\)
    with approximation factor \(1 - \varepsilon\),
    whose size is bounded by a polynomial in \(n\), i.e.,
    \(\fc(\oprob_n, 1 - \varepsilon) \leq \poly(n)\).
  \item \emph{fully polynomial-size linear programming
      relaxation scheme (FPSRS)}
    if
    for every \(\varepsilon > 0\)
    there is an LP formulation of \(\oprob_{n}\)
    with approximation factor \(1 - \varepsilon\),
    whose size is bounded by
    a polynomial in \(n\) and \(1 / \varepsilon \), i.e.,
    \(\fc(\oprob_n, 1 - \varepsilon) \leq \poly(n, 1 / \varepsilon)\)
    for every
    \(\varepsilon > 0\) and \(n\).
  \end{enumerate}
\end{definition}
Thus both PSRS and FPSRS require a polynomial \(p\)
with \(\fc(\oprob_{n}, 1 - \varepsilon) \leq p(n)\).
The difference between PSRSs and FPSRSs
is that an FPSRS requires \(p\)
to depend polynomially on \(1 / \varepsilon\) as well, whereas for 
a PSRS the polynomial \(p\) is allowed to depend arbitrarily on
\(1/\varepsilon\).

For example, it is known that the maximum knapsack problem admits a PSRS
as shown in \cite{bienstock2008approximate}, 
however it is not known whether it also admits an FPSRS.
The matching problem has the following folklore PSRS:

\begin{exa}[PSRS for matching]
  \label{ex:PSRS-matching}
  We use the standard realization of the matching problem
  via the matching polytope \(\match{n}\),
  i.e., represent matchings by their characteristic vectors.
  We claim that the standard realization with
  the following linear program
  is an LP formulation of size \(O(n^{1 / (1 - \rho)})\)
  for the matching problem with approximation factor \(\rho\),
  for \(0 < \rho < 1\).
  \begin{align*}
    x(\delta(v)) &\leq 1 & \forall v &\in [n] \\
    x(E[U]) &\leq \frac{\card{U} - 1}{2 \rho}
    &\forall U &\subseteq [n],
    \card{U} \text{ odd},  \size{U} < \frac{1}{1 - \rho} \\
    x &\geq 0.
  \end{align*}
  Let \(K_{n}\) be the polytope defined by these inequalities.
  For the claim, it is enough to prove
  \(\match{n} \subseteq K_{n} \subseteq \rho^{-1} \match{n}\).
  The first inequality is obvious, and for the second one,
  we need to prove that for every \(x \in K_{n}\)
  we have
  \(x(E[U]) \leq \frac{\card{U} - 1}{2 \rho}\)
  for \(U \subseteq [n]\) odd
  and \(\size{U} \geq \frac{1}{1 - \rho}\).
  This inequality follows by summing up
  the inequalities \(x(\delta(v)) \leq 1\)
  for \(v \in U\) and using \(x \geq 0\):
  \begin{equation*}
    x(E[U]) \leq \frac{1}{2} \sum_{v \in U} x(\delta(v))
    \leq \frac{\size{U}}{2} \leq \frac{\size{U} - 1}{2 \rho}.
  \end{equation*}
\end{exa}

\subsubsection{Slack matrix and factorization}
\label{sec:slack-matrix}

The main tool for lower bounding the formulation complexity
is via the nonnegative rank of the \emph{slack matrix}.

\begin{definition}[Slack matrix]
  Given an approximation problem
  \(\oprob^{*} = (\oprob, \mathcal{F}^{*})\)
  of a maximization problem \(\oprob = (\mathcal{S}, \mathcal{F})\),
  the \emph{slack matrix \(S\) of \(\oprob^{*}\)} is
  the \(\mathcal{F} \times \mathcal{S}\) matrix
  where \(S_{f, s} \coloneqq f^{*} - f(s)\) for all \(f,s\).
\end{definition}

The \emph{nonnegative rank \(\nnegrk S\)}
of a nonnegative matrix \(S\)
is the smallest nonnegative integer \(r\)
such that \(M = \sum_{i \in [r]} M_{i}\)
is a sum of \(r\) nonnegative rank-\(1\) matrices \(M_{i}\).
Yannakakis's factorization theorem identifies extension complexity of
a polytope with the nonnegative rank of any of its slack matrices (see
\cite{Yannakakis88,Yannakakis91}).  The theorem extends naturally to
the notion of \emph{formulation complexity}:

\begin{theorem}[{\cite[Theorem~3.3]{BSZ2014._lp_sdp}}]
  \label{thm:YannaApprox}
  Let \(\oprob^{*}\) be an approximation problem
  with slack matrix \(S\).
  Then \(-1 + \nnegrk S \leq \fc(\oprob^{*}) \leq \nnegrk S\). 
\end{theorem}
Thus, for a lower bound on formulation complexity,
it suffices to lower
bound the nonnegative rank of the slack matrix of the problem of
interest.
We refer the interested reader also to \cite{Pashkovich12,bfps2012} for
a similar theorem using the classical polyhedral pairs
approach.
Actually, formulation complexity is equal to
a modified version of nonnegative rank,
showing that it is completely determined by the slack matrix.
However, for our purposes here it will be sufficient to
consider the nonnegative rank of the slack matrix.

We will establish the non-existence of an FPSRS for the matching
problem by providing a lower bound for \(\fc(\matching{n}, \rho)\)
for suitably chosen \(\rho\).
For this, we consider the approximation problem
\(\matching{n}^{*}_{\varepsilon}\)
with \(0 < \varepsilon \leq 1\),
where the underlying maximization problem is the subproblem of
\(\matching{n}\)
with objective functions restricted to the \(f_{U}\)
for complete graphs \(K_{U}\)
for odd sets \(U \subseteq [n]\)
and feasible solutions only the perfect matchings \(M\).
In particular, we assume \(n\)
is even.  The intuition is that every maximal matching \(M\)
of a graph \(G\)
arises as the restriction of any perfect matching \(M'\),
obtained by extending \(M\),
and hence this restriction does not alter the maximum value of
functions.  The restriction to complete graphs on odd sets is
motivated by the LP formulation given by the matching polytope, where
the overwhelming majority of facets correspond to such graphs.

We choose approximation guarantees
slightly weaker than coming from the
approximation factor \(1 - \varepsilon / (n - 1)\), in order to
simplify the later analysis:
\begin{equation*}
  f^{*}_{K_{U}} \coloneqq \frac{\card{U} - 1 + \varepsilon}{2}
  \geq \frac{\card{U} - 1}{2 (1 - \varepsilon / (n - 1))},
\end{equation*}
where the inequality is equivalent to
\(\size{U} \leq n - \varepsilon\),
which clearly holds. With this choice the slack entries
depend only on the number of crossing edges:
\begin{equation*}
  S_{M, U}^{+ \varepsilon} \coloneqq
  \frac{\card{U} - 1 + \varepsilon}{2} - \size{E[U] \cap M}
  =
  \frac{\card{\delta(U) \cap M} - 1 + \varepsilon}{2},
\end{equation*}
whereas for an approximation factor \(\rho\),
the slack entries would also depend on the size of \(U\):
\[S_{M,U}^\rho \coloneqq
  \frac{\card{U} - 1}{2 \rho} - \size{E[U] \cap M}
= \frac{1}{2} \left(
  \card{\delta(U) \cap M} - 1 + \frac{1 - \rho}{\rho} (\card{U}-1)
\right),\]
which would complicate the analysis.

\subsection{Lower bounds on the nonnegative rank via common information}
\label{sec:lower-bounds-nonn}

We further extend the common information-based framework that was introduced in
\cite{BP2013commInfo} and expanded in \cite{BJLP2013}. The underlying
approach is based on the sampling framework introduced in
\cite{braverman2012information} and implicitly related to previous lower bounding techniques 
given in \cite{BarYossef2004702}. We bound the log nonnegative rank
from below by common information, an information-theoretic quantity
that was introduced in \cite{wyner1975common}. We recall the basic
framework here and apply it to the matching problem in
Section~\ref{sec:lower-bounds-extens}. 

\begin{defn}[Common information]
  \label{def:common-information}
  Let \(\VAR A, \VAR B\) be random variables,
  and \(\VAREVENT Z\) be a conditional.
  The \emph{common information} of \(\VAR A, \VAR B\) given
  \(\VAREVENT Z\)
  is the quantity
  \begin{equation}
    \label{eq:common-information}
    \commonInfo[\VAREVENT Z]{\VAR A}{\VAR B} \coloneqq
    \inf_{\substack{\VAR \Pi\colon \VAR  A \perp \VAR  B \mid \VAR \Pi \\
        \VAR \Pi \perp \VAREVENT Z \mid \VAR A, \VAR B}}
    \mutualInfo[\VAREVENT{Z}]{\VAR{A}, \VAR{B}}{\VAR{\Pi}},
  \end{equation}
  where the infimum is taken over all
  random variables \(\VAR \Pi\) in all extensions of the probability
  space,
  so that 
  \begin{enumerate}
  \item \(\VAR  A\) and \(\VAR  B\) are conditionally independent
    given \(\VAR \Pi\), and
  \item \(\VAREVENT  Z\) and \(\VAR \Pi\) are conditionally independent
    given \(\VAR  A\) and \(\VAR  B\).
  \end{enumerate}
We refer to \(\VAR \Pi\) as \emph{seed} whenever it satisfies the
above properties.
\end{defn}
Note that when the condition includes random variables,
expectation should be automatically taken over all of them
when computing the mutual information.
In particular, common information is \emph{not} a function of
the random variables in its condition.

The conditional independence of \(\VAREVENT  Z\) and \(\VAR \Pi\) is to ensure that no
information about \(\VAR \Pi\) should be leaked from \(\VAREVENT Z\). We will link
common information to nonnegative matrices, by reinterpreting the latter
as probability distributions over two random variables \(\VAR  A\) and \(\VAR  B\). 

\begin{defn}[Induced distribution]
\label{def:infrk}
  Let \(M \in \R_+^{m \times n}\) be a nonnegative matrix. The \emph{induced distribution}
 on \(\VAR  A,\VAR  B\) with
 \(\range(\VAR{A}) = [m]\) and \(\range(\VAR{B}) = [n]\) via \(M\)
 is given by
  \begin{equation*}
    \probability{\VAR  A = a, \VAR  B = b} =
    \frac{M(a, b)}{\sum_{x,y} M(x,y)}
  \end{equation*}
  for every row \(a\) and column \(b\). Slightly abusing notation, we
  write \((\VAR  A,\VAR  B) \sim M\).
\end{defn}

For convenience we define the \emph{common information of \(M\) conditioned on \(\VAREVENT  Z\)} as
  \begin{equation}
    \label{eq:def-infrk}
    \infrk[\VAREVENT Z]{M} \coloneqq \commonInfo[\VAREVENT Z]{\VAR
      A}{\VAR B},
  \end{equation}
where \((\VAR  A,\VAR  B) \sim M\). It can be shown that common information is a lower bound on the log of
the nonnegative rank.

\begin{lemma}[\cite{BP2013commInfo}]
  \label{lem:nonnegrank-via-IC}
  Every nonnegative factorization of a nonnegative matrix \(M\)
  induces a seed with range of size of the number of summands
  in the factorization.
  In particular, \(\log \nnegrk M \geq \infrk[\VAREVENT Z]{M}\)
  for any condition \(\VAREVENT Z\).
\end{lemma}
Even though it is not needed in the sequel,
for the sake of completeness,
we briefly recall the seed \(\VAR{\Pi}\) arising from a nonnegative matrix
factorization \(M = \sum_{i \in [r]} u_{i}^{\intercal} v_{i}\) of \(M\).
We start from a probability space \(\Omega\) containing
the random row \(\VAR{A}\), the random column \(\VAR{B}\),
and possibly other random variables,
needed to define \(\VAREVENT{Z}\).
Given an outcome of \(\Omega\),
we choose \(\VAR{\Pi} \in [r]\) with conditional distribution
\begin{equation*}
  \probability[\VAR{A}, \VAR{B}]{\VAR{\Pi} = i} =
  \frac{u_{i} (\VAR{A}) \cdot v_{i}(\VAR{B})}{M_{\VAR{A}, \VAR{B}}}
  .
\end{equation*}
This choice ensures the conditional independence requirements
to make \(\VAR{\Pi}\) a seed:
the independence of \(\VAR{\Pi}\) and \(\VAREVENT{Z}\)
given \(\VAR{A}, \VAR{B}\),
as well as the independence of \(\VAR{A}\) and \(\VAR{B}\)
given \(\VAR{\Pi}\).

Clearly, \(r \geq \entropy{\VAR{\Pi}} \geq
\mutualInfo[\VAREVENT{Z}]{\VAR{A}, \VAR{B}}{\VAR{\Pi}}\),
which by taking infimum leads to
\(\log \nnegrk M \geq \infrk[\VAREVENT Z]{M}\).

\section{Formulation complexity of the matching problem}
\label{sec:lower-bounds-extens}

We will now show how to bound the common information of the slack
matrix of the matching problem, which leads to a lower bound on the
nonnegative rank of the slack matrix
and hence the formulation
complexity of the matching problem via Theorem~\ref{thm:YannaApprox}
and Lemma~\ref{lem:nonnegrank-via-IC}. 

\begin{theorem}
  \label{th:approx-matching}
  Let \(0 < \varepsilon < 1\) be fixed and \(n\) even.
  Then \(\fc(\matching{n}^{*}_{\varepsilon}) = 2^{\Theta(n)}\).
  In particular,
  the formulation complexity of the matching problem
  with approximation factor \(1 - \varepsilon / n\)
  is
  \(\fc(\matching{n}, 1 - \varepsilon / n) = 2^{\Theta(n)}\).
  Therefore the matching problem does not admit
  an FPSRS.
\end{theorem}

Establishing Theorem~\ref{th:approx-matching} reduces to showing that 
\(\commonInfo[\VAREVENT{Z}]{\VAR{M}}{\VAR{U}} = \Omega(n)\)
for \((\VAR{M}, \VAR{U}) \sim S^{+ \varepsilon}\)
under a suitable condition \(\VAREVENT{Z}\) via Lemma~\ref{lem:nonnegrank-via-IC}  and
Theorem~\ref{thm:YannaApprox}.

\subsection{Preparation for the proof: probabilistic setup}
\label{sec:preparation-proof}

\begin{figure}[htb]
  \small
  \centering
  \includegraphics{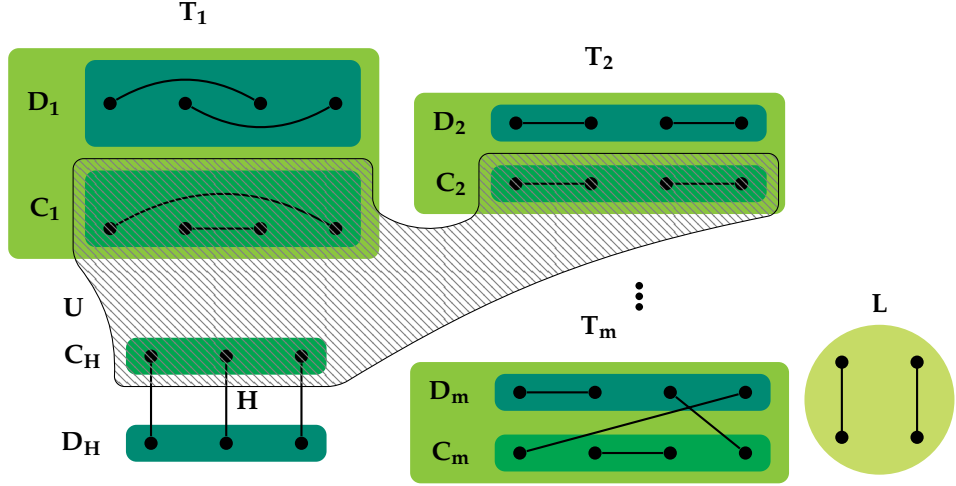}
  \caption{\label{fig:partition-Kn}%
    The condition \(\EVENT{E}\) together with a
    partition of \(K_{n}\) to reduce formulation complexity to
    fixed-size small chunks.}
\end{figure}

We need some preparations to introduce \(\VAREVENT{Z}\).
We partition the vertices of \(K_{n}\)
in a manner similar to but simpler than
in \cite{rothvoss2013matching},
see Figure~\ref{fig:partition-Kn}.
The purpose is to break up the graph into small chunks,
in order to amplify the complexity of the chunks.

First, we choose a \(3\)-matching \(\VAR{H}\)
between two disjoint \(3\)-element subsets
\(\VAR{C_{H}}\) and \(\VAR{D_{H}}\).
The intention is to consider only pairs
\((\VAR{M}, \VAR{U})\)
with \(\VAR{H} \subseteq \VAR{M}\)
and \(\delta(\VAR{U}) \cap \VAR{M} = \VAR H\),
together with \(\VAR{C_{H}} \subseteq \VAR{U}\)
and \(\VAR{U} \cap \VAR{D_{H}} = \emptyset\).
We partition the vertices not covered by \(\VAR{H}\)
into equal-sized chunks \(\VAR{T_{1}}, \dotsc, \VAR{T_{m}}\)
of even size \(2 (k - 3)\),
where \(k \geq 7\) is a fixed integer.
This might not be possible, and some vertices might be left out,
therefore we add a remainder
chunk \(\VAR{L}\) (which might be empty) of size \(l < 2 (k - 3)\).
In particular, \(n - 6 = 2 (k - 3) m + l\).
We partition every \(\VAR{T_{i}}\) into
a pair of \((k-3)\)-element sets \(\VAR{C_{i}}, \VAR{D_{i}}\).
We will denote by \(\VAR{T}\) the collection of
\(\VAR{C_{1}}, \VAR{D_{1}}, \dotsc, \VAR{C_{m}}, \VAR{D_{m}},
\VAR{C_{H}}, \VAR{D_{H}}, \VAR L\).
Let \(\VAR{T}\) and \(\VAR{H}\) be jointly uniformly distributed
independently of \(\VAR{M}, \VAR{U}\).

Second, we need a collection \(\VAR N\) of \(m\) mutually independent
random fair coins \(\VAR{N_{1}}, \dotsc, \VAR{N_{m}}\),
which are also independent of the random variables introduced before.
Finally, we introduce the events \(\EVENT{E}_{0}\) and \(\EVENT{E}\)
formalizing the above restrictions of \(\VAR{M}, \VAR{U}\)
imposed by \(\VAR{T}\):
  \begin{align*}
    \EVENT{E}_{0} &\coloneqq
    \left\{
      \begin{aligned}
        \VAR{U} \cap \VAR{T_{i}} &\in
        \{\emptyset, \VAR{C_{i}}\},
        &
        \VAR{C_{H}} \subseteq
        \VAR{U} &\subseteq
        \VAR{C_{H}} \cup \bigcup_{i \in [m]} \VAR{C_{i}}
        \\
        \VAR{H} &\subseteq \VAR{M},  &
        \VAR{M} &\subseteq \VAR{H} \cup E[\VAR{L}]
        \cup \bigcup_{i \in [m]} E[\VAR{T_{i}}]
      \end{aligned}
    \right.
    \\
    \EVENT{E} &\coloneqq \EVENT{E}_{0} \wedge
    \begin{cases}
      \VAR{U} \cap \VAR{T_{i}} = \emptyset, & \text{if } \VAR{N_{i}} = 0, \\
      \delta(\VAR{C_{i}}) \cap \VAR{M} = \emptyset,
      & \text{if } \VAR{N_{i}} = 1
      .
    \end{cases}
  \end{align*}
In particular, given \(\EVENT E_0\) we have \(\VAR{L} \cap \VAR{U} =
\emptyset\). 
Actually, the sole role of \(\VAR{L}\) is to collect the vertices
not fitting into the scheme.
The event \(\EVENT{E}\) ensures
\(\delta(\VAR{U}) \cap \VAR{M} = \VAR{H}\).

Now we are in the position to define \(\VAREVENT{Z}\)
as \(\VAR{T}, \VAR{H}, \VAR{N}, \EVENT{E}\),
and hence to formulate the exact lower bound on common information:

\begin{proposition}
  \label{prop:matching-technical}
  Let \(k \geq 7\) be a fixed odd number,
  \(0 \leq \varepsilon < 1 - 4/k\), and
  \(n = 2 (k - 3) m + l + 3\) for some \(l < k\).
  Consider the complete graph \(K_{n}\) on \(n\) vertices.
  Furthermore, let \(\VAR{M}\) be a random perfect matching
  and \(\VAR{U}\) a random subset of vertices of odd size, with
  \((\VAR M, \VAR U) \sim S^{+ \varepsilon}\).
  Then there exists a constant \(c_{k, \varepsilon}  > 0\)
  depending only on \(k\) and \(\varepsilon\),
  so that
  \begin{equation*}
    \commonInfo[\VAR{T}, \VAR{N}, \VAR{H}, \EVENT{E}]{\VAR{M}}{\VAR{U}}
    =
 \expectation(\VAR{T}, \VAR{N}, \VAR{H} \mid \EVENT{E})
 {\commonInfo[\VAR{T}, \VAR{N}, \VAR{H}, \EVENT{E}]
   {\VAR{M}}{\VAR{U}}}
    \geq c_{k, \varepsilon} m.
  \end{equation*}
\end{proposition}

With Proposition~\ref{prop:matching-technical}, we immediately obtain a proof
of Theorem~\ref{th:approx-matching}. 

\begin{proof}[Proof of Theorem~\ref{th:approx-matching}]
  
The upper bound \(2^{O(n)}\) follows from the LP formulation
given by the matching polytope.
For the lower bounds, note that the first bound
implies the second one,
as it is for a subproblem with weaker approximation guarantees:
Formally, mapping every feasible solution and objective function to
itself reduces the first problem to the second one,
hence
\(\fc(\matching{n}, 1 - \varepsilon / n) \geq
\fc(\matching{n}^{*}_{\varepsilon})\)
by \cite[Proposition~4.2]{BSZ2014._lp_sdp}.
The non-existence of an FPSRS follows as
for fixed \(\varepsilon > 0\)
we pick the approximation factor \(\rho = 1 - \varepsilon/n\),
which requires exponential size formulations.

Thus it remains to establish 
\(\fc(\matching{n}^{*}_{\varepsilon}) = 2^{\Omega(n)}\), which 
is obtained via Theorem~\ref{thm:YannaApprox}
by lower bounding the nonnegative rank of the slack matrix
\(S^{+ \varepsilon}\)
via Lemma~\ref{lem:nonnegrank-via-IC} and Proposition~\ref{prop:matching-technical}.
\end{proof}

Therefore it is sufficient to prove
Proposition~\ref{prop:matching-technical}, which we will do in the remainder of this section. The argument follows the framework in
\cite{BP2013commInfo}:
Let \(\VAR{\Pi}\) be a seed for \((\VAR{M}, \VAR{U})\).

\begin{enumerate}
\item \emph{Reduction to local case:} We first reduce the general case to the case \(m = 1\) and \(l =
  0\). This follows via a direct sum argument and the
  partitions from above serve the purpose of creating independent
  subproblems. 
\item \emph{Bound from local case:} We then analyze the case \(m =
  1\), \(l = 0\) for fixed \(k \geq 7\) and show that
\[\mutualInfo[\VAR{T}, \VAR{N}, \VAR{H}, \EVENT{E}]{\VAR{M}, \VAR{U}}{\VAR{\Pi}}
    \geq c_{k, \varepsilon},\]
via Pinsker's inequality (see Lemma~\ref{lem:pinsker}).
\end{enumerate}

\subsection{Reduction to the local case}
\label{sec:reduction-local-case}

We will now provide the reduction to the local case below and in
Section~\ref{sec:local-case} we provide the analysis of the local
case.

\begin{proof}[Proof of Proposition~\ref{prop:matching-technical}:
reduction to the local case with \(m = 1\) and \(l = 0\)]
Suppose that the statement of the proposition holds for \(m = 1\) and
\(l = 0\).

First, observe that the event \(\EVENT E\) ensures
that \(\delta(\VAR{U}) \cap \VAR{M} = \VAR{H}\). Thus, as
the probability of a pair \((\VAR M,\VAR U)\) depends only
on the number of
crossing edges, \((\VAR{M}, \VAR{U})\) is uniformly
distributed given
\(\EVENT E\).

The matching \(\VAR{M}\) decomposes into
\(\VAR{M_{i}} \coloneqq \VAR{M} \cap  E[\VAR{T_{i}}]\)
for \(i \in [m]\),
together with
\(\VAR{M_{L}} \coloneqq \VAR{M} \cap E[\VAR{L}]\) and \(\VAR{H}\).
Similarly, the set \(\VAR{U}\) decomposes as
\(\VAR{U} = \VAR{C_{H}} \cup \bigcup_{i \in [m]} \VAR{U_{i}}\)
with \(\VAR{U_{i}} \coloneqq \VAR{U} \cap \VAR{T_{i}}\).
The pairs \((\VAR{M_{i}}, \VAR{U_{i}})\) together with \((\VAR{M_L}, \emptyset)\)
are mutually independent,
therefore by the direct sum property
\begin{equation*}
  \mutualInfo[\VAR{T}, \VAR{N}, \VAR{H}, \EVENT{E}]{\VAR{M}, \VAR{U}}{\VAR{\Pi}}
  \geq
  \sum_{i \in [m]} \mutualInfo[\VAR{T}, \VAR{N}, \VAR{H}, \EVENT{E}]{\VAR{M_{i}}, \VAR{U_{i}}}{\VAR{\Pi}}
  \geq c_{k, \varepsilon} m,
\end{equation*}
where the last inequality is concluded from the local case as follows.

We prove that every summand is at least \(c_{k, \varepsilon}\)
via reduction to the local case.
Let us consider a fixed \(i \in [m]\),
and let us also fix \(\VAR{L}\) and \(\VAR{T_{j}}, \VAR{N_{j}}\) for
all \(j \neq i\).
This actually also fixes \(\VAR{T_{i}}\),
and hence the \(2k\)-element subset
\(V_{i} = \VAR{T_{i}} \cup \VAR{C_{H}} \cup \VAR{D_{H}}\).
Only the partition of \(V_{i}\) into \(\VAR{C_{i}}, \VAR{D_{i}},
\VAR{C_{H}}, \VAR{D_{H}}\) remains random. 
Let \(\EVENT{E}_{-i}\) be the event \(\EVENT{E}\)
with the restrictions for \(\VAR{U_{i}}\) and \(\VAR{M_{i}}\) omitted,
it ensures that all crossing edges lie inside \(V_{i}\).
Therefore, given \(\VAR{T_{j}}, \VAR{N_{j}}\) for \(j \neq i\)
and \(\EVENT{E}_{-i}\),
the distribution of
\(\VAR{M_{i}} \cup \VAR{H}, \VAR{U_{i}} \cup \VAR{C_{H}}, \VAR{\Pi},
\VAR{H}, \VAR{N_{i}}, \VAR{C_{i}}, \VAR{D_{i}},
\VAR{C_{H}}, \VAR{D_{H}}\)
on the complete graph on \(V_{i}\)
is exactly the one given in the proposition for the case \(m =
1\) and \(l = 0\). The events \(\EVENT{E}_{0}\) and \(\EVENT{E}\)
also have the same interpretation in the local and global case.

We invoke the proposition for the local case, which provides
\[\mutualInfo[\VAR{C_{i}}, \VAR{D_{i}}, \VAR{C_{H}}, \VAR{D_{H}},
\VAR{N_{i}}, \VAR{H},\EVENT E]{\VAR{M_{i}} \cup \VAR{H},
  \VAR{U_{i}} \cup \VAR{C_{H}}}{\VAR{\Pi}} \geq c_{k, \varepsilon}.\]
Here we have explicitly written out the partition of \(V_{i}\)
in the condition.
Replacing \(\VAR{M_{i}} \cup \VAR{H}\) with \(\VAR{M_{i}}\)
  \(\VAR{U_{i}} \cup \VAR{C_{H}}\) with \(\VAR{C_{H}}\)
does not alter the mutual information,
as given the condition,
\(\VAR{M_{i}} \cup \VAR{H}\) and \(\VAR{M_{i}}\)
determine each other,
and also
\(\VAR{U_{i}} \cup \VAR{C_{H}}\) and \(\VAR{C_{H}}\)
determine each other.
Unfixing the  \(\VAR{T_{j}}, \VAR{N_{j}}\) and \(\VAR{L}\),
and taking expectation over those
leads to
\(\mutualInfo[\VAR{T}, \VAR{N}, \VAR{H}, \EVENT{E}]
{\VAR{M_{i}}, \VAR{U_{i}}}{\VAR{\Pi}} \geq c_{k, \varepsilon}\),
as claimed.
\end{proof}

\subsection{Proof of Proposition~\ref{prop:matching-technical}: The local case}
\label{sec:local-case}

\begin{figure}[htb]
  \small
  \centering
  \includegraphics{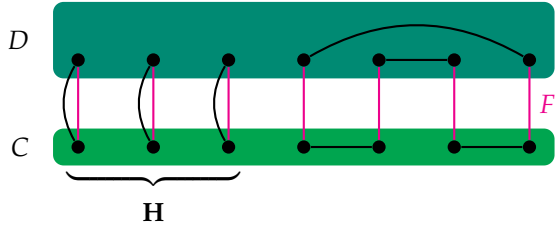}
  \caption{\label{fig:local-setup}%
    Condition in the local case.
    The blocks \(C\), \(D\) and a perfect matching \(F\) between them
    are fixed.}
\end{figure}

In the local case \(m = 1\), \(l = 0\),
we first adjust the setup,
as illustrated in Figure~\ref{fig:local-setup}.
We introduce some auxiliary random variables.
Let \(\VAR{C} \coloneqq \VAR{C_{1}} \cup \VAR{C_{H}}\)
and \(\VAR{D} \coloneqq \VAR{D_{1}} \cup \VAR{D_{H}}\).
Note that \(\VAR{C}, \VAR{D}\) and \(\VAR{H}\) are
uniformly distributed
(independently of \(\VAR{M}, \VAR{U}, \VAR{\Pi}, \VAR{N}\)),
and together determine
the \(\VAR{C_{1}}, \VAR{D_{1}}, \VAR{C_{H}}, \VAR{D_{H}}\).
Furthermore, we introduce \(\VAR{F}\) as
a uniformly random extension of \(\VAR{H}\)
into a full matching between \(\VAR{C}\) and \(\VAR{D}\),
depending only on \(\VAR{C}\), \(\VAR{D}\) and \(\VAR{H}\).
This independence ensures that adding it as condition
to the mutual information has no effect:
\begin{equation*}
 \begin{split}
  \mutualInfo[\VAR{T}, \VAR{H}, \VAR{N}, \EVENT{E}]
  {\VAR{M}, \VAR{U}}{\VAR{\Pi}}
  &
  =
  \mutualInfo[\VAR{T}, \VAR{H}, \VAR{F}, \VAR{N}, \EVENT{E}]
  {\VAR{M}, \VAR{U}}{\VAR{\Pi}}
  =
  \mutualInfo[\VAR{C}, \VAR{D}, \VAR{F}, \VAR{H}, \VAR{N}, \EVENT{E}]
  {\VAR{M}, \VAR{U}}{\VAR{\Pi}}
  \\
  &
  =
  \expectation(C \sim \VAR{C}, D \sim \VAR{D}, F \sim \VAR{F}
  \mid \EVENT{E})
  {\mutualInfo[\VAR{C} = C, \VAR{D} = D, \VAR{F} = F, \VAR{H},
    \VAR{N}, \EVENT{E}]{\VAR{M}, \VAR{U}}{\VAR{\Pi}}}
  .
 \end{split}
\end{equation*}
We show that the inner term is always at least \(c_{k, \varepsilon}\).
To this end, we fix \(C, D, F\),
and drop them from the condition to simplify notation.
For convenience, we rewrite the events \(\EVENT{E}_{0}\)
and \(\EVENT{E}\) in a simple form:
\begin{align*}
  \EVENT{E}_{0} &\coloneqq
  \VAR{U} \in \{C, C(\VAR{H})\} \land \VAR{H} \subseteq \VAR{M}
  &
  \EVENT{E} &\coloneqq
  \EVENT{E}_{0} \land
  \begin{cases}
    \VAR{U} = C(\VAR{H}), & \text{if } \VAR{N}=0 \\
    \delta(C) \cap \VAR{M} = \VAR{H}, & \text{if } \VAR{N}=1
    .
  \end{cases}
\end{align*}
Here and below
for a \(3\)-matching \(h \subseteq F\),
let \(C(h)\) denote the endpoints of the edges
of \(h\) lying in \(C\).
(I.e., \(C(\VAR{H}) = \VAR{C_{H}}\) but we shall use the notation
for other \(h\), too.)
As \(\VAR{N}\) consists of only one coin,
we simply identify it with \(\VAR{N}\).

To proceed, we will distinguish two types of pairs \((\pi , h)\):
the \emph{good pairs} are where the conditional distribution
\((\VAR M, \VAR U) \mid \VAR{\Pi} = \pi, H = h\) is close
to the distribution \((\VAR M, \VAR U) \mid H = h\)
with the condition on \(\VAR{\Pi}\) left out.
Therefore the contribution of good pairs to mutual information
is negligible.
The \emph{bad pairs} are where the two
distributions differ significantly,
and hence contribute much to mutual information.
As we will see in Section~\ref{sec:bound-prob-being},
there are not too many good pairs,
and this is the key to the proof.

We now state the exact definitions of goodness and badness,
taking also the conditions in the mutual information into account,
using a small positive number \(\delta > 0\) chosen later:
\begin{definition}
  \label{def:bad}
  A pair \((\pi, h)\) is \emph{\(\VAR{M}\)-good}
  if for all matchings
  \(m \supseteq h\)
  \begin{equation*}
    1 - \delta \leq
    \frac{\probability[\VAR{\Pi} = \pi, \VAR{H} = h,
      \VAR{N} = 0, \EVENT{E}]{\VAR{M} = m}}
    {\probability[\VAR{H} = h, \VAR{N} = 0, \EVENT{E}]
      {\VAR{M} = m}}
    \leq 1 + \delta.
  \end{equation*}
  Otherwise the pair is \emph{\(\VAR{M}\)-bad},
  denoted as \(\Mbad{\pi, h}\).
  Similarly,
  a pair \((\pi, h)\) is \emph{\(\VAR{U}\)-good}
  if for \(u = C(h)\) and \(u = C\)
  \begin{equation*}
    1 - \delta \leq
    \frac{\probability[\VAR{\Pi} = \pi, \VAR{H} = h,
      \VAR{N} = 1, \EVENT{E}]{\VAR{U} = u}}
    {\probability[\VAR{H} = h, \VAR{N} = 1, \EVENT{E}]
      {\VAR{U} = u}}
    \leq 1 + \delta.
  \end{equation*}
  Otherwise the pair is \emph{\(\VAR{U}\)-bad},
  denoted as \(\Ubad{\pi, h}\).

  The pair \((\pi, h)\) is \emph{good}
  if it is both \(\VAR{M}\)-good and \(\VAR{U}\)-good.
  It is denoted by \(\good{\pi, h}\).
  The pair is \emph{bad}, denoted as \(\bad{\pi, h}\)
  if it is not good.
\end{definition}

Now we reduce the proposition to estimating
the probability of bad pairs.

\begin{proof}[Proof of Proposition~\ref{prop:matching-technical}: the
  case \(m = 1\) and \(l = 0\)]
We will use Pinsker's inequality (Lemma~\ref{lem:pinsker})
to lower bound the mutual information induced by those \((\pi,h)\)
which are \(\VAR M\)-bad or \(\VAR U\)-bad (depending on the outcome
of the coin \(\VAR N\)). Therefore we rewrite the mutual information using relative entropy:
\begin{multline*}
  \mutualInfo[\VAR{H}, \VAR{N}, \EVENT{E}]
  {\VAR{M}, \VAR{U}}{\VAR{\Pi}}
  =
  \expectation(\VAR{\Pi}, \VAR{H}, \VAR{N} \mid \EVENT{E}){
    \relEntropy
    {\VAR{M}, \VAR{U} \mid \VAR{\Pi}, \VAR{H}, \VAR{N}, \EVENT{E}}
    {\VAR{M}, \VAR{U} \mid \VAR{H}, \VAR{N}, \EVENT{E}}}
  \\
  =
  \sum_{i \in \binSet} \probability[\EVENT{E}]{\VAR{N} = i} \cdot
  \expectation(\VAR{\Pi}, \VAR{H} \mid \VAR{N}=i, \EVENT{E}){
    \relEntropy
    {\VAR{M}, \VAR{U} \mid \VAR{\Pi}, \VAR{H}, \VAR{N} = i, \EVENT{E}}
    {\VAR{M}, \VAR{U} \mid \VAR{H}, \VAR{N} = i, \EVENT{E}}}
  .
\end{multline*}

By definition of badness, for \(\VAR M\)-bad pairs there is an \(m\)
where the probabilities differ by at least
\[\delta
\probability[\VAR{H} = h, \VAR{N} = 0, \EVENT{E}]{\VAR{M} = m}
= \delta \alpha,\]
where \(\alpha = 1 / (2k - 7)!!\),
the reciprocal of the
number of perfect matchings on \(2k-6\) nodes.
Similarly, for \(\VAR U\)-bad pairs,
the probabilities differ by at least
\[\delta
\probability[\VAR{H} = h, \VAR{N} = 1, \EVENT{E}]{\VAR{U} = u}
= \delta \frac{1}{2}.\]
Thus via Lemma~\ref{lem:pinsker} we have
\begin{multline*}
  \expectation(\VAR{\Pi}, \VAR{H} \mid \VAR{N}=0, \EVENT{E}){
    \relEntropy
    {\VAR{M}, \VAR{U} \mid \VAR{\Pi}, \VAR{H}, \VAR{N} = 0, \EVENT{E}}
    {\VAR{M}, \VAR{U} \mid \VAR{H}, \VAR{N} = 0, \EVENT{E}}}
  \\
  \geq
  \probability[\VAR{N} = 0, \EVENT{E}]{\Mbad{\VAR{\Pi}, \VAR{H}}}
  2 (\log e) (\delta \alpha)^{2}
  ,
\end{multline*}
and
\begin{multline*}
  \expectation(\VAR{\Pi}, \VAR{H} \mid \VAR{N}=1, \EVENT{E}){
    \relEntropy
    {\VAR{M}, \VAR{U} \mid \VAR{\Pi}, \VAR{H}, \VAR{N}=1, \EVENT{E}}
    {\VAR{M}, \VAR{U} \mid \VAR{H}, \VAR{N}=1, \EVENT{E}}}
  \\
  \geq
  \probability[\VAR{N} = 1, \EVENT{E}]{\Ubad{\VAR{\Pi}, \VAR{H}}}
  2 (\log e) \left( \frac{\delta}{2} \right)^{2}
  .
\end{multline*}

The main part of the proof is to lower bound the
probability of being bad. To simplify computations, we 
rewrite the probabilities appearing above in
a more manageable form:
\begin{align*}
 \probability[\VAR{N} = 0, \EVENT{E}]{\Mbad{\VAR{\Pi}, \VAR{H}}}
 &
 =
 \probability[\VAR{U} = C(\VAR{H}), \EVENT{E}_{0}]
 {\Mbad{\VAR{\Pi}, \VAR{H}}}
 \\
 &
 =
 \frac{\probability[\EVENT{E}_{0}]
   {\Mbad{\VAR{\Pi}, \VAR{H}}, \VAR{U} = C(\VAR{H})}}
 {\probability[\EVENT{E}_{0}]{\VAR{U} = C(\VAR{H})}}
 \\
 \probability[\VAR{N} = 1, \EVENT{E}]{\Ubad{\VAR{\Pi}, \VAR{H}}}
 &
 =
 \frac{\probability[\EVENT{E}_{0}]
   {\Ubad{\VAR{\Pi}, \VAR{H}}, \VAR{H} = \delta(C) \cap \VAR{M}}}
 {\probability[\EVENT{E}_{0}]{\VAR{H} = \delta(C) \cap \VAR{M}}}
 .
\end{align*}
We shall obtain a constant \(B_{k, \varepsilon} > 0\)
lower bounding the sum of the
probabilities of being \(\VAR M\)-bad or \(\VAR U\)-bad
\[\probability[\EVENT{E}_{0}]
{\Mbad{\VAR{\Pi}, \VAR{H}}, \VAR{U} = C(\VAR{H})}
+
\probability[\EVENT{E}_{0}]
{\Ubad{\VAR{\Pi}, \VAR{H}}, \VAR{H} = \delta(C) \cap \VAR{M}}
\geq B_{k, \varepsilon}.\]
Note that the
probabilities cannot be bounded separately as \emph{individually}
they can be \(0\).
Once obtained this bound then leads to
\begin{equation*}
 \begin{split}
  \mutualInfo[\VAR{H}, \VAR{N}, \EVENT{E}]
  {\VAR{M}, \VAR{U}}{\VAR{\Pi}}
  &
  \geq
  \probability[\EVENT{E}]{\VAR{N} = 0} \cdot
  \probability[\VAR{N} = 0, \EVENT{E}]{\Mbad{\VAR{\Pi}, \VAR{H}}}
  2 (\log e) (\delta \alpha)^{2}
  \\
  &
  +
  \probability[\EVENT{E}]{\VAR{N} = 1} \cdot
  \probability[\VAR{N} = 1, \EVENT{E}]{\Ubad{\VAR{\Pi}, \VAR{H}}}
  2 (\log e) \left( \frac{\delta}{2} \right)^{2}
  \\
  &
  \geq
  B_{k, \varepsilon} (2 \log e) \delta^{2} \min
  \left\{
    \frac{\probability[\EVENT{E}]{\VAR{N} = 0} \alpha^{2}}
    {\probability[\EVENT{E}_{0}]{\VAR{U} = C(\VAR{H})}},
    \frac{\probability[\EVENT{E}]{\VAR{N} = 1} / 4}
    {\probability[\EVENT{E}_{0}]{\delta(C) \cap \VAR{M} = \VAR{H}}}
  \right\}
  \eqqcolon c_{k, \varepsilon}
  .
 \end{split}
\end{equation*}
This is a positive constant depending only on \(k\) and
\(\varepsilon\) provided \(B_{k, \varepsilon} >0\), which we prove in the next
section.

\subsection{Bounding probability of being good}
\label{sec:bound-prob-being}

To obtain the claimed lower bound \(B_{k, \varepsilon} > 0\)
on the probability of being bad,
we investigate how much the good pairs contribute to the distribution
of \(\VAR{M}, \VAR{U}\).
We start by rewriting \emph{goodness} into a form
with less conditions on probabilities.
For any \(3\)-matching \(h\) between \(C\) and \(D\),
any perfect matching \(m \supseteq h\)
we have
\begin{align*}
  \probability[\VAR{\Pi} = \pi, \VAR{H} = h, \VAR{N} = 0,
  \EVENT{E}]{\VAR{M} = m}
  &
  =
  \probability[\VAR{\Pi} = \pi, \VAR{H} = h, \VAR{N} = 0,
  \VAR{U} = C(h), h \subseteq \VAR{M}]{\VAR{M} = m}
  \\
  &
  =
  \probability[\VAR{\Pi} = \pi,
  \VAR{U} = C(h), h \subseteq \VAR{M}]{\VAR{M} = m}
  \\
  &
  =
  \probability[\VAR{\Pi} = \pi, h \subseteq \VAR{M}]{\VAR{M} = m}
  =
  \frac{\probability[\VAR{\Pi} = \pi]{\VAR{M} = m}}
  {\probability[\VAR{\Pi} = \pi]{h \subseteq \VAR{M}}}
  ,
\end{align*}
by first expanding \(\EVENT{E}\),
then removing \(\VAR{H}, \VAR{N}\)
using their independence of \(\VAR{M}, \VAR{\Pi}, \VAR{U}\),
and finally removing \(\VAR{U}\),
as it is independent of \(\VAR{M}\) given \(\VAR{\Pi}\).
Similarly, for \(u \in \{C(h), C\}\)
\begin{equation*}
  \probability[\VAR{\Pi} = \pi, \VAR{H} = h, \VAR{N} = 1,
  \EVENT{E}]{\VAR{U} = u}
  =
  \frac{\probability[\VAR{\Pi} = \pi]{\VAR{U} = u}}
  {\probability[\VAR{\Pi} = \pi]{\VAR{U} \in \{C(h), C\}}}
  .
\end{equation*}
This is mostly useful for comparing probabilities for the various
values of \(m, u\).
Let \(  \beta \coloneqq
  \left(
    \frac{1 - \delta}{1 + \delta}
  \right)^{2}
\).
When \((\pi, h)\) is good, we obtain
\begin{align*}
  \sqrt{\beta}
  \probability[\VAR{\Pi} = \pi]{\VAR{U} = C}
  \leq
  \probability[\VAR{\Pi} = \pi]{\VAR{U} = C(h)}
  &
  \leq
  \frac{1}{\sqrt{\beta}}
  \probability[\VAR{\Pi} = \pi]{\VAR{U} = C}
  , & \text{ and}
  \\
  \sqrt{\beta}
  \probability[\VAR{\Pi} = \pi]{\VAR{M} = F}
  \leq
  \probability[\VAR{\Pi} = \pi]{\VAR{M} = m}
  &
  \leq
  \frac{1}{\sqrt{\beta}}
  \probability[\VAR{\Pi} = \pi]{\VAR{M} = F}
  ,
  &
  \text{for } h &\subseteq m.
\end{align*}
Let us fix \(\pi\), and let \(h_{1}, \dotsc, h_{N}\)
be all the \(3\)-submatchings of \(F\) with \((\pi, h_{i})\)
good.
Let \(u_{i} = C(h_{i})\) denote the vertices of \(h_{i}\) in \(C\).
We divide the \(h_{i}\) into two classes:
the first class consists of the \(h_{i}\)
having two common edges with every other \(h_{j}\),
i.e., the intersection \(u_{i} \cap u_{j}\) has \(2\) elements
for all \(j \neq i\).
Without loss of generality,
the first class is \(h_{1}, \dotsc, h_{\ell}\).
The second class consists of the \(h_{i}\)
which do not have two common edges with all the other \(h_{j}\).

The Erdős–Ko–Rado Theorem \cite[Theorem~2]{ErdosKoRado1961}
bounds the size of the first class by
\(\ell \leq k-2\) for \(k \geq 6\),
(slightly improving over the \(3k\) in the combinatorial part of
\cite[Lemma~8]{rothvoss2013matching}).
We consider matchings \(m\) satisfying
\(h_{i} = \delta(C) \cap m\) for an \(i\).
For \(i \leq \ell\), i.e., \(h_{i}\) is in the first class,
we estimate
\begin{equation*}
 \begin{split}
  \probability[\VAR{\Pi} = \pi]{\VAR{M} = F, \VAR{U} = C}
  &
  =
  \probability[\VAR{\Pi} = \pi]{\VAR{M} = F}
  \probability[\VAR{\Pi} = \pi]{\VAR{U} = C}
  \\
  &
  \geq
  \sqrt{\beta}
  \probability[\VAR{\Pi} = \pi]{\VAR{M} = m}
  \cdot
  \sqrt{\beta}
  \probability[\VAR{\Pi} = \pi]{\VAR{U} = u_{i}}
  \\
  &
  =
  \beta
  \probability[\VAR{\Pi} = \pi]{\VAR{M} = m, \VAR{U} = u_{i}}
  .
 \end{split}
\end{equation*}
Taking the average over all \(m\), we obtain
\begin{equation}
  \label{eq:prob-good-k}
  \probability[\VAR{\Pi} = \pi]{\VAR{M} = F, \VAR{U} = C}
  \geq
  \frac{\beta}{A}
  \probability[\VAR{\Pi} = \pi]{h_{i} = \delta(C) \cap \VAR{M},
    \VAR{U} = u_{i}}
  ,
  \qquad i \leq \ell
  ,
\end{equation}
with
\(A \coloneqq (k-4)!!^{2}\) being the number of matchings \(m\)
with \(\delta(C) \cap m = h_{i}\).

For \(i > \ell\) there is a \(j\)
such that \(\card{u_{i} \cap u_{j}} \leq 1\).
Thus there is an edge \(e \in E[u_i]\)
not incident with any vertices of \(u_{j}\).
We replace the two edges \(e_{1}, e_{2}\) in \(F\)
at the endpoints of \(e\) by the two edges
\(e\) and \(f\) connecting their endpoints
in \(C\) and \(D\), respectively,
see Figure~\ref{fig:cross-over-1-crossing}.
\begin{figure}[htb]
  \small
  \centering
  \includegraphics{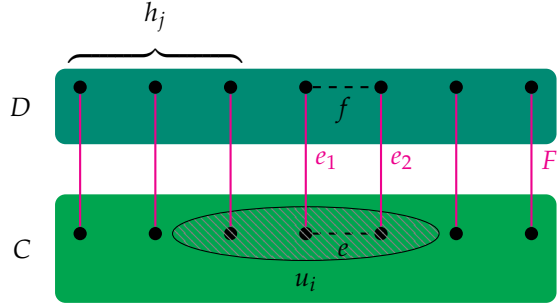}
  \caption{\label{fig:cross-over-1-crossing}%
    Constructing the matching \(m'\) by replacing \(e_{1}\),
    \(e_{2}\) in \(F\) by the dashed edges:
    \(e\) in \(u_{i}\) and its counter part \(f\) in \(D\).}
\end{figure}
Let \(m' \coloneqq F \setminus \{e_{1}, e_{2}\} \cup \{e, f\}\)
be the new matching, it contains \(e\).
Note that
\(\size{\delta(u_{i}) \cap m'} = 1\)
and
\begin{equation*}
 \begin{split}
  \probability[\VAR{\Pi} = \pi]{\VAR{M} = m', \VAR{U} = u_{i}}
  &
  =
  \probability[\VAR{\Pi} = \pi]{\VAR{M} = m'}
  \probability[\VAR{\Pi} = \pi]{\VAR{U} = u_{i}}
  \\
  &
  \geq
  \sqrt{\beta}
  \probability[\VAR{\Pi} = \pi]{\VAR{M} = F}
  \probability[\VAR{\Pi} = \pi]{\VAR{U} = u_{i}}
  \\
  &
  \geq
  \beta
  \probability[\VAR{\Pi} = \pi]{\VAR{M} = m}
  \probability[\VAR{\Pi} = \pi]{\VAR{U} = u_{i}}
  \\
  &
  =
  \beta
  \probability[\VAR{\Pi} = \pi]{\VAR{M} = m, \VAR{U} = u_{i}}
  .
 \end{split}
\end{equation*}
Taking the average over all \(m\) provides
\begin{equation}
  \label{eq:prob-good-1}
 \begin{split}
  \probability[\VAR{\Pi} = \pi]{
    \size{\delta(\VAR{U}) \cap \VAR{M}} = 1,
    \size{\VAR{M} \setminus F} = 2,
    \VAR{U} = u_{i}}
  &
  \geq
  \probability[\VAR{\Pi} = \pi]{\VAR{M} = m', \VAR{U} = u_{i}}
  \\
  &
  \geq
  \frac{\beta}{A}
  \probability[\VAR{\Pi} = \pi]{h_{i} = \delta(C) \cap \VAR{M},
    \VAR{U} = u_{i}}
  .
 \end{split}
\end{equation}
Now we add \(\VAR{H}\) to \eqref{eq:prob-good-k} and \eqref{eq:prob-good-1},
using its independence of the variables involved there, we have
\begin{multline}
  \label{eq:1}
  \frac{\beta}{A}
  \probability[\VAR{\Pi} = \pi]{\VAR{H} = \delta(C) \cap \VAR{M},
    \VAR{U} = C(\VAR{H}), \VAR{H} = h_{i}}
  \\
  \leq
  \begin{cases}
    \probability[\VAR{\Pi} = \pi]{\VAR{M} = F, \VAR{U} = C}
    / \binom{k}{3}
    ,
    & \text{for } i \leq \ell
    ,
    \\
    \probability[\VAR{\Pi} = \pi]
    {\size{\delta(\VAR{U}) \cap \VAR{M}} = 1,
    \size{\VAR{M} \setminus F} = 2,
    \VAR{H} = h_{i}, \VAR{U} = C(\VAR{H})}
    ,
    & \text{for } i > \ell
  .
  \end{cases}
\end{multline}
We sum up over all \(i\) to obtain
\begin{multline}
  \label{eq:2}
  \frac{\beta}{A}
  \probability[\VAR{\Pi} = \pi]{\VAR{H} = \delta(C) \cap \VAR{M},
    \VAR{U} = C(\VAR{H}), \good{\VAR{\Pi}, \VAR{H}}}
  \\
  \leq
  \frac{\ell}{\binom{k}{3}}
  \probability[\VAR{\Pi} = \pi]{\VAR{M} = F, \VAR{U} = C}
  +
  \probability[\VAR{\Pi} = \pi]
  {\size{\delta(\VAR{U}) \cap \VAR{M}} = 1,
    \size{\VAR{M} \setminus F} = 2,
    \VAR{U} = C(\VAR{H})}
  .
\end{multline}
We take expectation over \(\pi \sim \VAR \Pi\).
Recall that \(\ell \leq k - 2\),
hence
\begin{equation}
  \label{eq:3}
 \begin{split}
  \frac{\beta}{A}
  &
  \left(\strut
  \probability{\VAR{H} = \delta(C) \cap \VAR{M},
    \VAR{U} = C(\VAR{H})}
  -
  \probability{\VAR{H} = \delta(C) \cap \VAR{M},
    \VAR{U} = C(\VAR{H}), \bad{\VAR{\Pi}, \VAR{H}}}
  \strut\right)
  \\
  &
  =
  \frac{\beta}{A}
  \probability{\VAR{H} = \delta(C) \cap \VAR{M},
    \VAR{U} = C(\VAR{H}), \good{\VAR{\Pi}, \VAR{H}}}
  \\
  &
  \leq
  \frac{k-2}{\binom{k}{3}}
  \probability{\VAR{M} = F, \VAR{U} = C}
  +
  \probability{\size{\delta(\VAR{U}) \cap \VAR{M}} = 1,
    \size{\VAR{M} \setminus F} = 2,
    \VAR{U} = C(\VAR{H})}
  .
 \end{split}
\end{equation}
We compute the values of the various probabilities from
\(S^{+\varepsilon}\).
The probability of a fixed pair \((M, U)\)
is proportional to their slack value, i.e.,
\begin{equation*}
  \probability{\VAR{M} = M, \VAR{U} = U}
  =
  (\size{\delta(U) \cap M} - 1 + \varepsilon) \gamma,
\end{equation*}
for some positive constant \(\gamma\) depending only on \(k\)
and \(\varepsilon\).
In particular,
\begin{equation*}
  \probability{\VAR{M} = F, \VAR{U} = C}
  =
  (k - 1 + \varepsilon) \gamma.
\end{equation*}
(We do not need the exact value of \(\gamma\), but actually
\(\gamma =
\left[
  \sum_{M, U} (\size{\delta(\VAR{U}) \cap \VAR{M}} - 1 + \varepsilon)
\right]^{-1}
=
[(k/2 - 1 + \varepsilon) \cdot (2k - 1)!! \cdot 2^{2k - 1}]^{-1}\),
using that there are \((2k-1)!! 2^{2k - 1}\) pairs \((M, U)\)
altogether,
and \(\binom{2k}{2}\) edges of \(K_{2k}\).
Each edge belongs to \(\delta(U) \cap M\)
for exactly an \(1 / [2 (2 k - 1)]\) part of the pairs \((M, U)\).)

For the other events,
it is easier to compute the probability conditioned on \(\VAR{H}\):
\begin{equation*}
  \probability{\VAR{H} = \delta(C) \cap \VAR{M}, \VAR{U} = C(\VAR{H})}
  =
  \expectation(\VAR{H}){
    \probability[\VAR{H}]
    {\VAR{H} = \delta(C) \cap \VAR{M}, \VAR{U}= C(\VAR{H})}}
  = (2 + \varepsilon) \gamma A.
\end{equation*}
Recall that \(A\) is the number of pairs satisfying the event
for \(\VAR{H}\) fixed.
Here we have used that \((\VAR{M}, \VAR{U})\)
is independent of \(\VAR{H}\).

The probability of the third event is determined similarly.
Note that for any fixed \(3\)-element \(\VAR{U}\),
there are \(3\) matchings \(m\) with
\(\size{\delta(\VAR{U}) \cap m} = 1\)
and \(\size{m \setminus F} = 2\),
actually arising by an analogous construction
depicted in Figure~\ref{fig:cross-over-1-crossing}.
\begin{equation*}
  \probability{\size{\delta(\VAR{U}) \cap \VAR{M}} = 1,
    \size{\VAR{M} \setminus F} = 2,
    \VAR{U} = C(\VAR{H})}
  =
  \varepsilon \gamma \cdot 3
  .
\end{equation*}
Substituting the probabilities back into our formula,
we obtain
\begin{equation*}
  \probability{\VAR{H} = \delta(C) \cap \VAR{M},
    \VAR{U} = C(\VAR{H}), \bad{\VAR{\Pi}, \VAR{H}}}
  \geq
  (2 + \varepsilon) \gamma A
  -
  \left(
    \frac{6 (k - 1 + \varepsilon)}{k (k - 1)} \gamma
    +
    \varepsilon \gamma \cdot 3
  \right)
  \frac{A}{\beta}
  .
\end{equation*}

Finally, we add back the event \(\EVENT{E}_{0}\) as a condition.
Recall that \(\varepsilon \leq 1 - 4/k\), hence
\begin{multline*}
  \probability[\EVENT{E}_{0}]{\VAR{H} = \delta(C) \cap \VAR{M},
    \VAR{U} = C(\VAR{H}), \bad{\VAR{\Pi}, \VAR{H}}}
  \\
  \geq
  \left(
    2 + \varepsilon
    -
    \left(
      \frac{6 (k - 1 + \varepsilon)}{k (k - 1)}
      +
      \varepsilon \cdot 3
    \right)
    \frac{1}{\beta}
  \right)
  \frac{\gamma A}
  {\probability{\EVENT{E}_{0}}}
  \\
  =
  \left(
    2
    -
    \left(
      \frac{6 (k - 1 + \varepsilon)}{k (k - 1)}
      +
      (3 - \beta) \varepsilon
    \right)
    \frac{1}{\beta}
  \right)
  \frac{\gamma A}
  {\probability{\EVENT{E}_{0}}}
  \\
  \geq
  \left(
    2
    -
    \left(
      \frac{6 (k - 1 + (1 - 4/k) )}{k (k - 1)}
      +
      (3 - \beta) (1 - 4/k)
    \right)
    \frac{1}{\beta}
  \right)
  \frac{\gamma A}
  {\probability{\EVENT{E}_{0}}}
  \\
  =
  \left(
    2 - (3 k - 4) (1 - \beta)
    - \frac{6 (k - 4)}{k (k - 1)}
  \right)
  \frac{\gamma A}
  {k \beta \probability{\EVENT{E}_{0}}}
  .
\end{multline*}
The last expression has a positive lower bound
\(B_{k, \varepsilon}\) depending only on \(k\) and \(\varepsilon\).
(Note that \(\probability{\EVENT{E}_{0}} / \gamma\)
involves \(\varepsilon\).)
To this end, we choose
\(\delta\) sufficiently close to \(0\),
ensuring that \(\beta\) is sufficiently close to \(1\),
making
the last parenthesized expression positive.
The number \(A\) depends just on \(k\).

All in all,
\(\probability[\EVENT{E}_{0}]{\VAR{H} = \delta(C) \cap \VAR{M},
    \VAR{U} = C(\VAR{H}), \bad{\VAR{\Pi}, \VAR{H}}}
\geq B_{k, \varepsilon} > 0\).
Finally,
\begin{multline*}
  \probability[\EVENT{E}_{0}]{\VAR{U} =  C(\VAR{H}),
    \Mbad{\VAR{\Pi}, \VAR{H}}}
  +
  \probability[\EVENT{E}_{0}]{\VAR{H} = \delta(C) \cap \VAR{M},
    \Ubad{\VAR{\Pi}, \VAR{H}}}
  \\
  \geq
  \probability[\EVENT{E}_{0}]{\VAR{H} = \delta(C) \cap \VAR{M},
    \VAR{U} = C(\VAR{H}), \bad{\VAR{\Pi}, \VAR{H}}}
  \geq
  B_{k, \varepsilon}
  .
\end{multline*}  
\end{proof}

\section{Concluding remarks}
\label{sec:concluding-remarks}
We believe that the information-theoretic framework is general enough
to also be able to reproduce the results in \cite{chan2013approximate}
and potentially improving on the strength of the lower bound. Also,
recently in \cite{LRS14} it has been shown that the results in
\cite{chan2013approximate} also extend to the case of semidefinite
programming formulation. However, it is open whether the
semidefinite formulation complexity of the matching problem is
exponential and a careful modification of the presented approach might
shed some light on this question.

\section*{Acknowledgements}
\label{sec:acknowledgements} 
The authors would like to thank Thomas Rothvoß for the helpful comments. Research was partially supported
by NSF grants CMMI-1300144 and CCF-1415496.

\bibliographystyle{abbrvnat}
\bibliography{bibliography}

\end{document}